# Plasmonic elliptical nanoholes for chiroptical analysis and enantioselective optical trapping

Zhan-Hong Lin,[a] Jiwei Zhang [a,b] and Jer-Shing Huang [a,c,d,e*]

**A simple yet effective achiral platform using elliptical nanoholes for chiroptical analysis is demonstrated. Under linearly polarized excitation, an elliptical nanohole in a thin gold film can generate a localized chiral optical field for chiroptical analysis and simultaneously serve as a near-field optical trap to capture dielectric and plasmonic nanospheres. In particular, the trapping potential is enantioselective for dielectric nanospheres, i.e., the hole traps or repels the dielectric nanoparticles depending on the sample chirality. For plasmonic nanospheres, the trapping potential well is much deeper than that for dielectric particles, rendering the enantioselectivity less pronounced. This platform is suitable for chiral analysis with nanoparticle-based solid-state extraction and pre-concentration. Compared to plasmonic chiroptical sensing using chiral structures or circularly polarized light, elliptical nanoholes are a simple and effective platform, which is expected to have a relatively low background because chiroptical noise from the structure or chiral species outside the nanohole is minimized. The usage of linearly polarized excitation also makes the platform easily compatible with a commercial optical microscope.**

## Introduction

Chiral light-matter interaction is of great research interest because it reveals the structural details of chiral matters,[1-4] such as the secondary structure of proteins and peptides.[5] However, due to the mismatch between the pitch length of the light helix and the size of molecular chiral domains,[2, 3] the molecule cannot "feel" the helicity of light, resulting in a weak chiroptical response. Apart from using standing waves generated by far-field interference to boost the optical field,[4, 6, 7] another way to enhance the chiroptical responses is to tailor the optical near field at the length scale comparable to the molecular size using rationally designed plasmonic nanostructures.

Resonant plasmonic nanostructures allow the manipulation and concentration of the electromagnetic field at the optical frequency in the near-field region and thereby promote the light-matter interaction.[8] Chiral plasmonic nanostructures and chiral metasurfaces have been applied to enhance chiral light-matter interaction in chiroptical spectroscopy,[4, 6, 9-14] chiral sensing,[15-18], chiral plasmonic nanostructures fabrication,[19] and photothermal circular dichroism (CD).[20, 21] However, under the excitation of circularly polarized light (CPL), the chiroptical response from the chiral nanostructures may introduce unwanted background that hinders the detection of the signal from the targeted chiral species and reduces the sensitivity. To solve this problem, CPL excitation on achiral plasmonic nanostructures and metasurfaces has been proposed and demonstrated.[13, 22-24] In this case, the structure is achiral but the excitation is still chiral. The distortion of the polarization state of CPL due to optical components, such as dichroic mirrors, is inevitable, rendering the experiment difficult and the interpretation of the data complex.[25] Alternatively, linearly polarized light (LPL) has been used to excite chiral plasmonic nanostructures and metasurfaces for chiral analysis.[9, 12, 26] The chirality of the chiral near field is determined by the nanostructures rather than the excitation.[12] This limits flexibility and tunability as the structures are fixed once fabricated. To get rid of the false chiroptical response due to the use of chiral substrate or chiral light, the ultimate solution is to use linearly polarized excitation and achiral nanostructures.[27, 28] This requires a careful design of the nanostructures so that LPL can be locally converted into an optical near field for specific chiral light-matter interaction. In this work, we proposed using elliptical nanoholes in an extended metal film for chiroptical analysis because they are achiral and can effectively convert LPL into local CPL.

Apart from generating the chiral optical field, another important issue for chiroptical analysis is that the targeted chiral analytes must be physically brought into the "hot zone" to interact with the specifically engineered chiral optical fields. This can be achieved by exploiting plasmonic near-field optical trapping.[29-33] Compared to conventional optical traps based on far-field optics, plasmonic near-field optical traps provide a larger field gradient under similar excitation power. Plasmonic nanoholes on a metallic film are powerful optical traps and have been used to capture extremely small dielectric nanoparticles or proteins.[34-38] To obtain higher field enhancement and localization, double nanoholes and bowtie apertures with sharp

[a.] *Leibniz Institute of Photonic Technology, Albert-Einstein Straße 9, 07745 Jena, Germany. E-mail: jer-shing.huang@leibniz-ipht.de*
[b.] *MOE Key Laboratory of Material Physics and Chemistry under Extraordinary Conditions, and Shaanxi Key Laboratory of Optical Information Technology, School of Physical Science and Technology, Northwestern Polytechnical University, Xi'an 710129, China*
[c.] *Abbe Center of Photonics, Friedrich-Schiller University Jena, Jena, Germany*
[d.] *Research Center for Applied Sciences, Academia Sinica, 128 Sec. 2, Academia Road, Nankang District, 11529 Taipei, Taiwan*
[e.] *Department of Electrophysics, National Chiao Tung University, 1001 University Road, 30010 Hsinchu, Taiwan*
† Electronic Supplementary Information (ESI) available: [details of any supplementary information available should be included here]. See DOI: 10.1039/x0xx00000x

tips have been designed for plasmonic optical trapping.[38-42] There are several advantages of using nanoholes as plasmonic optical traps. First, nanohole-based optical traps can stably capture dielectric particles for a very long time because the existence of the captured nanoparticles leads to a change of the local refractive index in the hole and thereby induces the so-called self-induced back-action, which stabilizes the particle in the hole.[34-36] Secondly, optical trapping using nanoholes in an extended metal film suffers less from the disturbance of thermal convection flow than solitary plasmonic structures because the extended metal film is an effective heat sink. Thirdly, using nanoholes in the metallic film is beneficial for optical detection in transmission mode because the nanohole enforces the light-matter interaction and the background is highly suppressed by the film. By smartly arranging the nanoholes into periodic arrays, it is possible to exploit extraordinary transmission to further enhance the transmission signal.[43, 44]

In this work, we propose using elliptical nanoholes in a gold film (Figure 1a) as a simple chiroptical analysis platform that combines the capabilities of optical trapping and plasmonic chiral sensing. We theoretically demonstrate that under LSP excitation, a chiral optical near field can be generated in the nanohole, which simultaneously provides a steep and trapping potential for 20-nm dielectric and metallic nanospheres. This platform avoids using circularly polarized illumination and thus is compatible with typical optical microscopes. The usage of achiral illumination and achiral aperture structures also avoids unwanted chiroptical backgrounds.

## Results and Discussion

### Generation of Chiral Optical Field in the Nanohole by Linearly Polarized Excitation

As the first step, we perform numerical simulations to find out the resonant radius of a circular nanohole in a gold film coated on a glass substrate. This serves as a starting point in the search for the optimal geometry of the elliptical hole. As indicated in Figure 1a, the radii of the nanohole along the $x$- and $y$-axis are $R_x$ and $R_y$, respectively. The whole platform is embedded in water and a laser source of 830 nm is used for excitation. The polarization angle of the normally incident plane wave from the glass half-space is defined by the in-plane angle $\theta$ with respect to the $+x$ axis. The numerical simulations were performed using the finite-difference time-domain method (FDTD Lumerical Solutions 8.18). Details of the simulations can be found in the Methods. The polarization angle of the excitation is fixed at $\theta = 45°$ to ensure equal electric field amplitude in the $x$- ($E_{x0}$) and $y$- direction ($E_{y0}$). The radius ($R$) of the circular nanohole is scanned from 60 to 140 nm with a step of 5 nm and the electric field intensity enhancement $\hat{I} = E_x^2 + E_y^2 + E_z^2$ is recorded at the "top center" of the nanohole, which is at the same height as the water-gold interface as marked by the red dot in Figure 1a. Here, $E_x$, $E_y$ and $E_z$ are the enhanced electric field components along the $x$-, $y$-, and $z$- direction, respectively. Figure 1b shows the field intensity enhancement ($\hat{I}$) as a function of the radius of the circular nanohole. A clear maximum is observed at $R = 105$ nm, which serves as a reference radius for the search of the dimensions of the elliptical nanoholes in the next step.

The next step is to find the best ellipticity of the elliptical nanohole that gives the maximal field enhancement and the highest degree of circular polarization under linearly polarized illumination. The degree of circular polarization $D_{\text{CPL}}$ is defined as $D_{\text{CPL}} = \frac{S_3}{S_0} = 2\frac{\langle E_x(t) E_y(t) \sin(\delta_x - \delta_y) \rangle}{\hat{I}}$. We use the field at the top center of the nanohole for the optimization of $D_{\text{CPL}}$. Here, $S_0$ and $S_3$ are the zeroth and third Stokes parameters, respectively.[45, 46] $E_x(t)$ and $E_y(t)$ are the time-variant electric field amplitudes along the $x$- and $y$-axis, respectively. $\delta_x$ and $\delta_y$ are the corresponding phases of the electric field components. The operation $\langle \rangle$ takes the time-averaged value. The value of $D_{\text{CPL}}$ lies between $-1$ and $+1$, which correspond to purely left- and right-handed CPL, respectively. Starting from the resonant radius ($R = 105$ nm) found in Figure 1b, we scanned $R_x$ and $R_y$ and evaluate the $\hat{I}$ and $D_{\text{CPL}}$ of the field in the elliptical nanohole. The radius $R_x$ is scanned from 50 to 105 nm and the radius $R_y$ is scanned from 105 to 205 nm with a step of 5 nm. As the radius changes, the resonance condition is detuned from the resonant wavelength and the phase of the corresponding field component gradually changes, leading to specific ellipticity of the local electric field in the nanohole. Once the phase shift between the $x$ and $y$ field component is around $\pi/2$ and the field amplitudes are the same, the local electric field is circularly polarized. In this case, $D_{\text{CPL}}$ is maximized and the elliptical nanohole functions as a nano quarter-wave plate that converts a linearly polarized far field to a circularly polarized near field. Since the nanohole will be used for optical trapping, we also map the field intensity enhancement $\hat{I}$ on the plane of $R_x$ and $R_y$. The goal is to find the $R_x$ and $R_y$ for the elliptical nanohole to provide the highest $\hat{I}$ and $D_{\text{CPL}}$. We found that using linearly polarized excitation at $\theta = 45°$, it is impossible to simultaneously obtain the highest $\hat{I}$ and $D_{\text{CPL}}$ (see Supporting

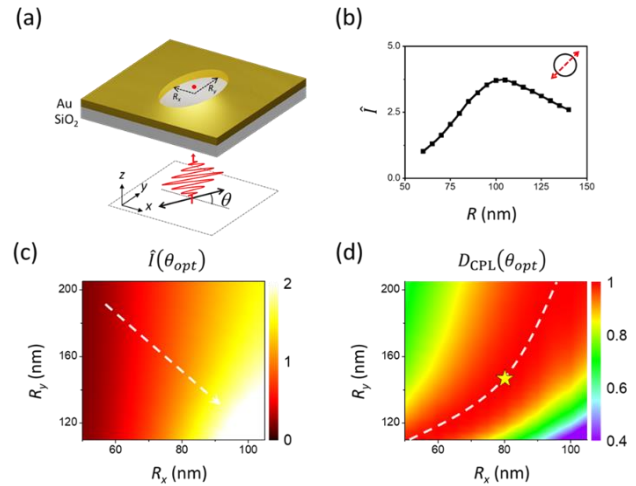

Figure 1. (a) Schematic of an elliptical nanohole in a thin gold film on a glass substrate for optical trapping and chiroptical analysis. The thickness of the gold film is set to be 100 nm. (b) The near-field intensity enhancement $\hat{I}$ as a function of $R$ of a circular nanohole. The inset shows the polarization state of the excitation light at $\theta = 45°$. (c) The near-field intensity enhancement $\hat{I}$, and (d) $D_{\text{CPL}}$ of the optical field in the elliptical nanohole as functions of $R_x$ and $R_y$. The optimal polarization angle ($\theta_{opt}$) of the excitation has been tuned for each structure to make sure $E_x = E_y$. The star marks the radii of the optimal elliptical nanohole used in this work.

Information for details). Therefore, we allow for the adjustment of the polarization angle $\theta$ in order to reach an equal field amplitude. This offers an additional degree of freedom to tune the ratio of the excitation field in the $x$- and $y$- direction ($\frac{E_{x0}}{E_{y0}}$) and balances the near electric field along the $x$- and $y$-axis in the elliptical nanohole, i.e., $E_x = E_y$. The adjustment of $\theta$ is optimized individually for each nanohole. Figures 1c and 1d show the $\hat{I}$ and $D_{\mathrm{CPL}}$ as functions of $R_x$ and $R_y$ obtained with the optimized excitation polarization. The $\hat{I}$ increases with increasing $R_x$ and decreasing $R_y$, as indicated by the dashed arrow in Figure 1c. A trajectory of $D_{\mathrm{CPL}} = 1$ on the plane of $R_x$ and $R_y$ is marked with a white dashed curve in Figure 1d. By comparing the trends of $\hat{I}$ and $D_{\mathrm{CPL}}$, we identify the best nanoholes geometry at radii $R_x = 80$ nm and $R_y = 145$ nm under a linearly polarized excitation at $\theta = 58.53°$. With this radius, the elliptical nanohole is expected to provide a circularly polarized field with the maximal near-field intensity enhancement.

In the following, we evaluate the optical chirality (OC) $C = -\frac{\varepsilon_0 \omega}{2} \mathrm{Im}(\tilde{\mathbf{E}}^* \cdot \tilde{\mathbf{B}})$ of the engineered optical near field because OC has been used for the evaluation of various chiral light-matter interaction, especially CD.[2,3] Here, $\varepsilon_0$ is the permittivity of vacuum and $\omega$ is the angular frequency. $\tilde{\mathbf{E}}$ and $\tilde{\mathbf{B}}$ are the complex local electric and magnetic fields. Rationally designed nanostructures can enhance the OC of the near field. The OC enhancement (OCE) factor is typically defined as the ratio of the OC of the near field to that of a perfect circularly polarized plane wave in the far-field,[27] $\hat{C} = \frac{C^\pm}{|C^\pm_{\mathrm{CPL}}|}$. Here, $C_{\mathrm{CPL}}$ is the OC of a far-field circularly polarized plane wave and "+ (−)" denotes the left- (right-) handedness. Figures 2 summarizes the simulated $\hat{I}$, $D_{\mathrm{CPL}}$, and $\hat{C}$ distribution inside the resonant circular nanohole ($R = 105$ nm) and the optimal elliptical nanoholes ($R_x = 80$ nm and $R_y = 145$ nm) under various excitation polarizations. Figures 2a and 2b show the results from a resonant circular nanohole under left-handed CPL and LPL ($\theta = 45°$) illumination. Figures 2c and 2d show the simulated field properties of the optimal elliptical hole under LPL illumination at the optimal polarization angle for the generation of left and right-handed optical near field inside the hole. Illuminating a resonant circular nanohole with CPL is a straightforward way to perform chiral analysis. As shown in Figure 2a, this method gives satisfactory field enhancement and OCE. However, this method is less favorable because the chiral optical field is not confined in the nanohole, leading to a relatively low signal enhancement and high noise from unwanted chiral light-matter interaction outside the nanohole. Moreover, using CPL as an illumination source makes it difficult to combine with a commercialized microscope because the optics inside the microscope can distort the circular polarization and lead to artifacts.[25,47] Using LPL (Figure 2b) to excite a circular nanohole, as expected, does not generate a chiral optical field inside the nanohole, despite the high field enhancement.[27] This field is not useful for chiroptical analysis. Differently, illuminating an optimally designed elliptical nanohole ($R_x = 80$ nm and $R_y = 145$ nm) with LPL at optimized polarization angles ($\theta = 58.53°$ and $121.47°$) effectively generate circularly polarized near field with significantly high field enhancement inside the elliptical nanohole, as shown in Figures 2c and d. The generated $\hat{I}$, $D_{\mathrm{CPL}}$, and $\hat{C}$ are comparable to those provided by a circular nanohole under the illumination of CPL. Using achiral elliptical nanohole with linearly polarized excitation effectively should generate the lowest chiral background noise from the nanostructures or species outside the hole. LPL illumination also makes the elliptical nanohole platform easily compatible with a commercial microscope. Moreover, the handedness of the circularly polarized near field in the elliptical nanohole can be easily flipped by changing the polarization of the illumination, i.e., from $\theta = 58.53°$ for left- to $\theta = 121.47°$ for right-handed near field. This further facilitates the quantitative measurement of CD.

**Optical Trapping Force Exerted on the Trapped Nanosphere**

Since the chiral near field is highly localized inside the elliptical nanohole, the ability to bring targets into the hot zone is important for plasmon-enhanced chiroptical sensing. In the following, we show that the optical near field in the elliptical nanohole can provide sufficient optical force to capture a 20-nm achiral dielectric or metal nanosphere into the nanohole. The optimal elliptical nanohole selected in the previous section is used here, i.e., an elliptical nanohole with radii of $R_x = 80$ nm and $R_y = 145$ nm excited by $58.53°$ linearly polarized excitation at 830 nm. We performed FDTD simulations to calculate the optical trapping force exerted on dielectric and plasmonic nanospheres located at 1 nm above the gold film. A 20-nm nanosphere made of gold (refractive index = 0.16138+5.1558$i$) or polystyrene (PS, refractive index = 1.575) was used as the trapping target. We modeled these nanospheres because they are commercially available and can be easily surface-functionalized to capture target analytes for the purposes of solid-state extraction and pre-concentration. [48]

The transverse optical force on a chiral object can be evaluated using the following equation, [32, 33, 49]

$$F_{tr} = \frac{\mathrm{Re}(\alpha_{ee})}{4}\nabla|E|^2 + \mathrm{Im}(\alpha_{em})\frac{1}{2}\nabla Im(E \cdot H^*), \qquad (1)$$

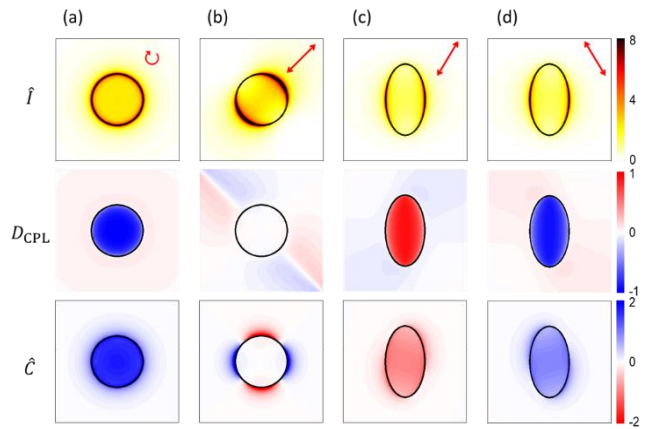

Figure 2. Distributions of $\hat{I}$, $D_{\mathrm{CPL}}$ and $\hat{C}$ inside the resonant circular ((a) and (b)) and the optimal elliptical nanohole ((c) and (d)). The radius of the resonant circular nanohole is 105 nm. The radii of the elliptical nanohole are $R_x = 80$ nm and $R_y = 145$ nm. The excitation polarization is indicated in the insets of the first row. The exact in-plane angles of the linear polarization in (b), (c), and (d) are $\theta = 45°$, $\theta = 58.53°$, and $\theta = 121.47°$, respectively.

where $\alpha_{ee}$ and $\alpha_{em}$ are the effective electric and electromagnetic polarizability of the chiral specimen, respectively. These polarizabilities can further be expressed as a function of the size, optical property, and chirality of the target, as $\alpha_{ee} = 4\pi r^3 \frac{(\varepsilon_r - \varepsilon_m)(\mu_r + 2) - \kappa^2}{(\varepsilon_r + 2\varepsilon_m)(\mu_r + 2) - \kappa^2}$ and $\alpha_{em} = -12\pi r^3 \frac{i\kappa\sqrt{\mu_0 \varepsilon_0}}{(\varepsilon_r + 2\varepsilon_m)(\mu_r + 2) - \kappa^2}$, where $r$ is the radius of nanosphere, $\varepsilon_r$ and $\mu_r$ ($\mu_r = 1$) are the relative permittivity and permeability of the particle, respectively, $\varepsilon_m$ is the relative permittivity of the medium, and $\kappa$ is the chirality parameter determining the degree of the handedness of the chiral material.[50] The first term in Equation 1 is the conventional gradient optical force ($F_{grad}$), and the second term is the gradient chiral force ($F_{chiral}$). Note that Equation 1 is valid under the assumption that the specimen is much smaller than the wavelength of light, *i.e.*, within the Rayleigh regime, and does not perturb the optical field. This condition is reasonable for the 20-nm nanospheres considered in this work. With the trapping force, the trapping potential can be obtained by integrating the in-plane force components along the corresponding direction, *e.g.*, $U_{xy} = \int F_x \cdot dx + \int F_y \cdot dy$.

In the following, we calculate the trapping potential due to the forces exerted on achiral and chiral nanospheres made of PS and gold. For achiral nanospheres, $\kappa$ is zero. Therefore, the second term of Equation 1, $F_{chiral}$, is zero. In this case, only the $F_{grad}$ is contributing to the trapping force. Taking the permittivity and permeability of PS and gold at 830 nm, the Re($\alpha_{ee}$) is calculated to be $1.31579 \times 10^{-35}$ and $1.36783 \times 10^{-34}$ for the PS and gold nanosphere, respectively. Gold nanosphere experiences about one order of magnitude larger gradient force than the PS nanosphere. This difference is due to the much larger electric polarizability of the gold nanosphere than that of the PS. The contour maps of the trapping potential for a 20-nm gold and PS nanosphere are shown in the upper and lower panel of Figure 3a, respectively. Both maps clearly show a saddle-like potential surface and two trapping spots inside the elliptical nanohole. The trapping potential for a 20-nm dielectric nanoparticle reaches 14 K$_B$T/W, which is sufficient to overcome the Brownian motion due to thermal energy.[35, 51] The presence of the nanosphere in the nanohole does not significantly perturb the optical field inside the nanohole. Figures 3b shows distributions of $\hat{I}$, $D_{CPL}$, and $\hat{C}$ of the near field calculated with the presence of the nanosphere. They are almost identical to those shown in Figure 2c. This means that the resonance condition of the elliptical nanohole is not significantly perturbed by the presence of the nanosphere. We note that this condition might change if more particles are trapped inside the nanohole.

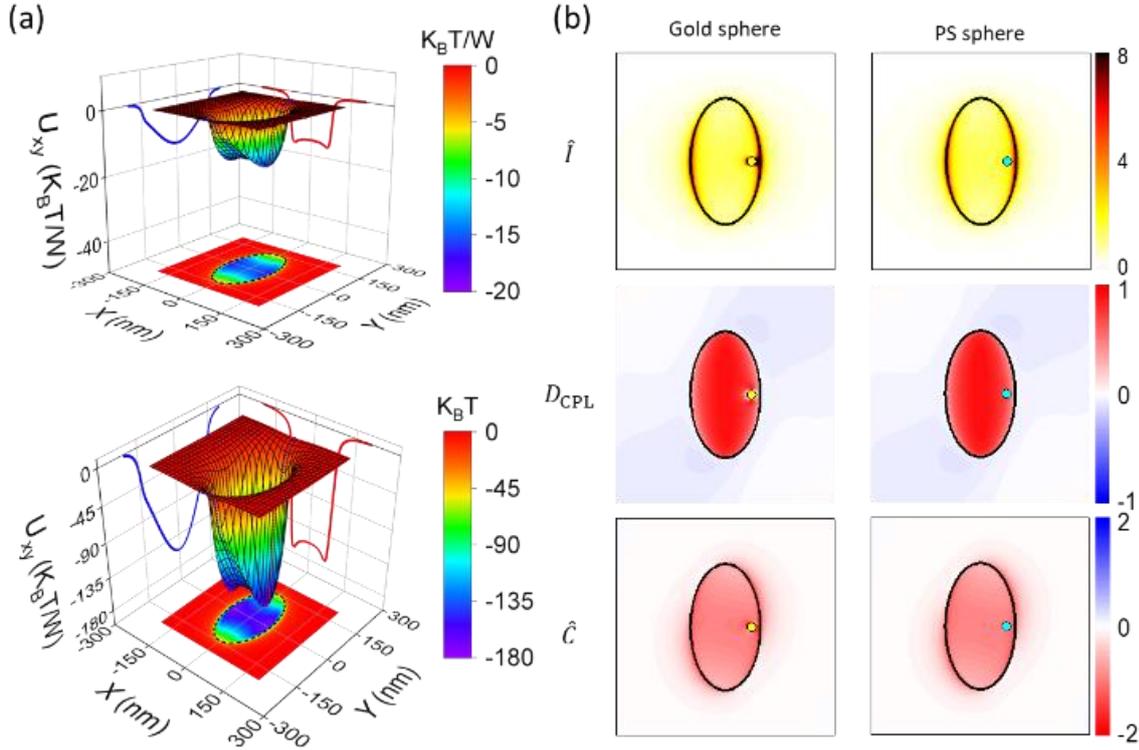

Figure 3. (a) Trapping potential contour maps of the in-plane optical force ($F_x$ and $F_y$) exerted on a PS (top panel) and gold (bottom panel) nanosphere (diameter = 20 nm). The red and blue curves on the x-z and y-z projection plane mark the potential value. The black dashed lines on the x-y projection plane mark the boundary of the elliptical nanohole. (b) Simulated distributions of $\hat{I}$, $D_{CPL}$ and $\hat{C}$ with the presence of the trapped gold (left panel) and PS (right panel) nanosphere at the right trapping site in the elliptical nanohole.

## Enantioselective Optical Trapping Force Exerted on the Trapped Nanosphere

Finally, we examine the enantioselectivity of the trapping force exerted on the nanoparticles. This analysis is useful for experiments, where the surface of the nanosphere is functionalized to capture chiral targets. The effective chirality of the nanoparticle covered with chiral targets can be described by introducing a non-zero $\kappa$, of which a $+(-)$ sign of the value indicates S(R) enantiomers.[32, 33] In this case, the electromagnetic polarizability $\alpha_{em}$ would have a finite value and the second term in Equation 1 also contributes to the overall trapping force, making the proposed platform enantioselective. For a 20-nm PS nanosphere with $\kappa = \pm 1$, the small permittivity $\mu_r$ of the dielectric material results in a relatively large $\alpha_{em}$ and, therefore, a $F_{\text{chiral}}$ about twice as large as $F_{\text{grad}}$. This leads to completely different optical potentials, corresponding to trapping and repelling, for the two enantiomers with opposite chirality. As shown in Figure 4a, when $\kappa = +1$, the signs of $F_{\text{grad}}$ and $F_{\text{chiral}}$ are the same, which means that the two terms sum up to create a deeper trapping selectively trap dielectric nanospheres according to the chirality of the targets captured on the surface of the nanospheres. For metallic nanospheres (Figures 4c and 4d), the trapping potential is still chirality dependent. However, the enantioselectivity is no longer available because the relatively large permittivity of metal leads to a very large $\alpha_{ee}$ and small $\alpha_{em}$. As a result, $F_{\text{grad}}$ is the dominant force, which is almost two orders of magnitude larger than $F_{\text{chiral}}$ and the trapping potential is always negative (trapping), regardless of the chirality of targets captured on the surface of the gold nanosphere.

## Conclusions

In conclusion, we have theoretically studied a simple yet effective platform of elliptical nanoholes in a gold film for simultaneous optical trapping and chiroptical analysis. By carefully designing the aspect ratio of the elliptical nanohole, linearly polarized excitation can be effectively converted into an enhanced circularly polarized field inside the nanohole. We potential than that in Figure 3a. Differently, when $\kappa = -1$, $F_{\text{chiral}}$ is larger than $F_{\text{grad}}$ and in opposite direction with $F_{\text{grad}}$, meaning that a negative trapping potential is created in the nanohole to repel the nanosphere, as shown in Figure 4b. This chirality-dependent optical potential is very useful because it offers the opportunity to show that the trapping potential generated by the optimal elliptical nanohole is sufficient to capture a 20-nm nanosphere made of PS or gold. The trapping potential well for the gold nanosphere is about an order of magnitude deeper than that of a PS nanosphere of the same size. The presence of the trapped nanosphere does not significantly alter the resonance condition of the elliptical nanohole. We also show that the elliptical nanohole offers enantioselective trapping potential that can selectively trap or repel PS nanosphere depending on the chirality of the nanosphere. As for gold nanoparticles, the contribution of the gradient trapping force is much larger than that of the chirality dependent gradient chiral force. As a result, gold nanoparticles will always be trapped into the elliptical nanohole regardless of their chirality. The proposed platform utilizes achiral nanostructures and achiral illumination. Therefore, it avoids false chiral signals from the plasmonic structures and the molecules outside of the hot zone do not contribute to the signal. Using linear excitation also makes the proposed platform compatible with commercial optical microscopes. Combining with suitably functionalized nanoparticles, the optical trapping platform based on the elliptical nanohole is expected to provide high sensitivity and selectivity for chiral sensing. The nanoholes may also be designed into resonant two-dimensional arrays to exploit the effect of extraordinary transmission at a specific wavelength for further improvement of the sensitivity.

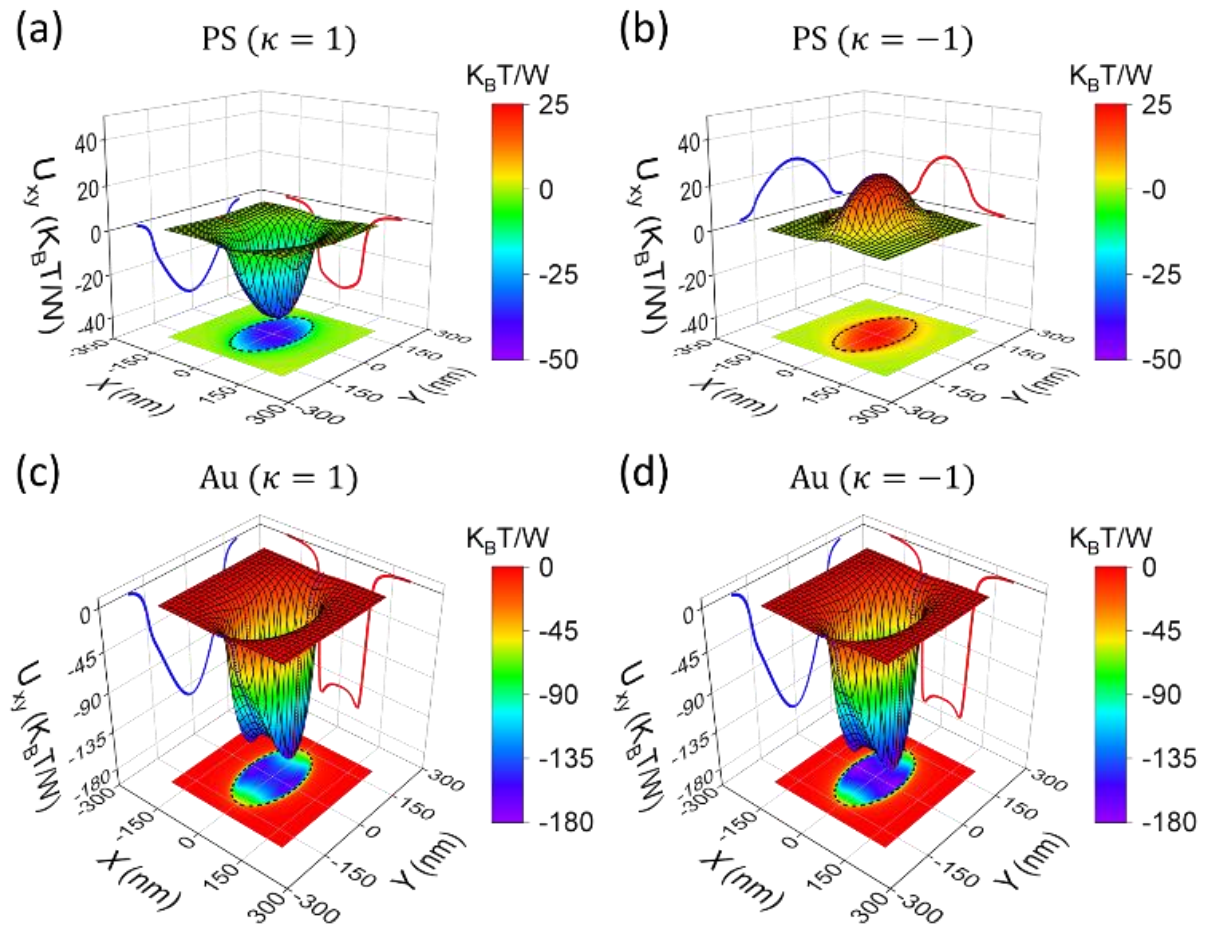

Figure 4. Trapping potential contour maps of the in-plane optical force ($F_x$ and $F_y$) exerted on a chiral (a, b) PS (top panel) and (c, d) gold (bottom panel) nanosphere (diameter = 20 nm). The chirality parameter $\kappa$ is $\pm 1$. The red and blue traces on the x-z and y-z projection planes are the cross-sectional potential cut by y = 0 and x = 0 planes. The black dashed lines on the x-y projection planes mark the boundaries of the elliptical nanohole.

## Methods

**Numerical Simulations.** For numerical simulations, three-dimensional finite-difference time-domain method (FDTD Solutions v8.18, Lumerical Solutions, Canada) was applied to calculate the optical response of the nanohole. The total-field scattered-field (TFSF) source was used to provide a linearly polarized light to excite the nanostructures. Circularly polarized plane waves were synthesized using two TFSF sources with orthogonal polarization states and a phase difference of $\pm\pi/2$. The substrate was set to be SiO$_2$ and the surrounding medium was set to be water. For all the simulations, the FDTD calculation region was set as 1.8 μm ($x$) × 1.8 μm ($y$) × 1.9 μm ($z$) with perfectly matched layers (PML, 12 layers) in all directions. The nanostructure was discretized with a 5 nm mesh step in a volume of 1.8 μm ($x$) × 1.8 μm ($y$) × 110 nm ($z$). The mesh step of 1 nm was used for the volume of 500 nm ($x$) × 500 nm ($y$) × 110 nm ($z$) that covers the nanohole. Such small mesh sizes can reasonably describe the curved surface of the nanohole structure and avoid the artifacts due to the non-physical geometrical features. For the simulations in Figure 1 in the main text, a point monitor was positioned at the top center of the nanohole to record the near-field optical response. To visualize the field distributions in Figures 2-4 in the main text, we recorded the near-field optical response on the $x-y$, $x-z$, and $y-z$ planes by using 2D field profile monitors. Besides, a 3D field profile monitor with a volume of 30 nm ($x$) × 30 nm ($y$) × 30 nm ($z$) covering the nanosphere was positioned in the same center as the nanosphere.

## Author Contributions

Z.L. and J.H. conceived of the presented idea. Z.L. performed the numerical simulations and analysed the data. All authors contributed to the final version of the manuscript. J.H. supervised the project.

## Conflicts of interest

There are no conflicts to declare.

## Acknowledgments

Financial supports from the Deutsche Forschungsgemeinschaft (CRC 1375 NOA, HU2626/3-1, HU2626/5-1) and the Ministry of Science and Technology of Taiwan (MOST-103-2113-M-007-004-MY3) are gratefully acknowledged. J. Z. acknowledges the support from Sino-German (CSC-DAAD) Postdoc Scholarship Program, 2018.etc.

–Electronic Supplementary Information –

# Plasmonic elliptical nanoholes for chiroptical analysis and enantioselective optical trapping

Zhan-Hong Lin,[1] Jiwei Zhang[1,2] and Jer-Shing Huang[1,3,4,5,*]

[1]*Leibniz Institute of Photonic Technology, Albert-Einstein Straße 9, 07745 Jena, Germany*

[2] *MOE Key Laboratory of Material Physics and Chemistry under Extraordinary Conditions, and Shaanxi Key Laboratory of Optical Information Technology, School of Physical Science and Technology, Northwestern Polytechnical University, Xi'an 710129, China*

[3]*Abbe Center of Photonics, Friedrich-Schiller University Jena, Jena, Germany*

[4]*Research Center for Applied Sciences, Academia Sinica, 128 Sec. 2, Academia Road, Nankang District, 11529 Taipei, Taiwan*

[5]*Department of Electrophysics, National Chiao Tung University, 1001 University Road, 30010 Hsinchu, Taiwan*

e-mail address: jer-shing.huang@leibniz-ipht.de

## S.1 Circularly Polarized Field Generation in the Nanohole by Linearly Polarized Excitation at θ=45°

The simulated $\hat{I}$ and $D_{\text{CPL}}$ under the linear excitation polarized at $\theta = 45°$ as functions of $R_x$ and $R_y$ have shown that the maximum near-field $\hat{I}$ and the $D_{\text{CPL}}$ cannot be simultaneously obtained. While small radii support large $\hat{I}$, high $D_{\text{CPL}}$ requires large radii, as indicated by the arrows in Figure S1a and b. The above problem stems from the fact that the elliptical nanohole cannot provide equal enhancement for the electric field components along the $x$- and $y$-axis. Consequently, the electric field components in $x$ and $y$ direction are not the same and the electric field inside the nanohole is elliptically polarized. To address this issue, the in-plane polarization of the linearly polarized illumination was optimized for every individual elliptical nanohole, as described in the main text.

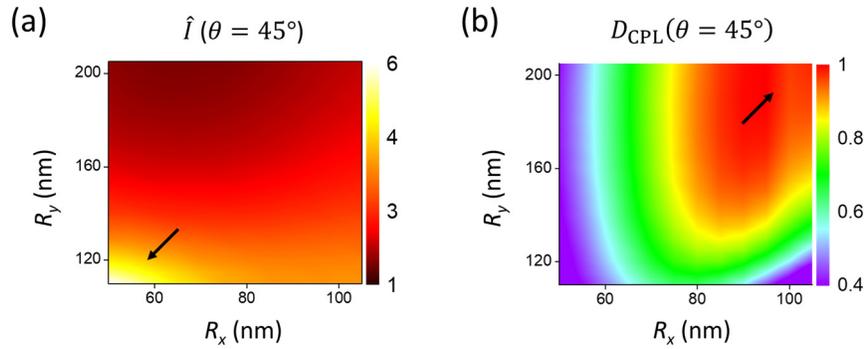

**Figure S1.** (a) Near-field intensity enhancement $\hat{I}$ and (b) $D_{\text{CPL}}$ of the field inside the elliptical nanohole verse $R_x$ and $R_y$ under a 45° polarized excitation.